\documentclass[12pt]{article}
\usepackage{graphicx}
\usepackage{epsfig,amsfonts,amssymb}
\usepackage{hyperref}
\usepackage{cite}
\input epsf.sty
\usepackage{cite}

\def\ZZZ{{\hbox{ Z\kern-1.6mm Z}}}
\def\RRR{{\hbox{ R\kern-2.4mm R}}}
\def\RRRR{{\hbox{ R\kern-4.1mm R}}}
\def\CCC{{\hbox{ C\kern-2.0mm C}}}
\def\zzz{{\hbox{z\kern-1mm z}}}

\newcommand{\qeq}{{\hbox{=\kern-2.3mm ? \kern.5mm }}}
\renewcommand{\qeq}{=}

\newcommand{\eps}{\epsilon}

\newcommand{\ra}{\rangle}
\newcommand{\la}{\langle}

\newcommand{\HH}{{\cal H}}

 \newcommand{\rrr}{\rangle\rangle}
 \renewcommand{\lll}{\langle\langle}

\newcommand{\wh}{\widehat}

\newcommand{\NN}{{\cal N}}

\newcommand{\be}{\begin{equation}}
\newcommand{\ee}{\end{equation}}
\newcommand{\ben}{\begin{eqnarray}\displaystyle}
\newcommand{\een}{\end{eqnarray}}

\newcommand{\refb}[1]{(\ref{#1})}

\newcommand{\sectiono}[1]{\section{#1}\setcounter{equation}{0}}

\newcommand{\eqref}{\refb}

\def\one{{\hbox{ 1\kern-.8mm l}}}
\def\zero{{\hbox{ 0\kern-1.5mm 0}}}

\begin{document}

\baselineskip 24pt

\begin{center}
{\Large \bf  State Operator Correspondence
and
Entanglement in AdS$_2$/CFT$_1$}

\end{center}

\vskip .6cm
\medskip

\vspace*{4.0ex}

\baselineskip=18pt

\centerline{\large \rm    Ashoke Sen}

\vspace*{4.0ex}

\centerline{\large \it $^a$Harish-Chandra Research Institute}
\centerline{\large \it  Chhatnag Road, Jhusi,
Allahabad 211019, India}

\vspace*{1.0ex}
\centerline{E-mail: sen@hri.res.in, 
ashokesen1999@gmail.com}

\vspace*{5.0ex}

\renewcommand{\check}{\bar }

\newcommand{\lan}{\langle\langle}
\newcommand{\ran}{\rangle\rangle}

\centerline{\bf Abstract} \bigskip

Since euclidean global AdS$_2$ space represented as
a strip has two boundaries, the state / operator
correspondence in the dual CFT$_1$ 
reduces to the standard
map from the operators acting on a single 
copy of the Hilbert space to states in the tensor product
of two copies of the Hilbert space. 
Using this picture we argue that the  
corresponding states in the dual string theory living
on AdS$_2\times {\rm K}$ are described by twisted version of the
Hartle-Hawking states,  the twists being generated by a
large unitary 
group of symmetries that this string theory must
possess.
This formalism makes natural the dual interpretation
of the black hole entropy, -- 
as the logarithm of the degeneracy of
ground states of the quantum mechanics describing the low
energy dynamics of the black hole, and
also as an entanglement entropy between the two copies of
the same quantum theory living
on the two boundaries of global AdS$_2$ separated by the
event horizon.

\vfill \eject

\baselineskip=18pt

\tableofcontents

\sectiono{Introduction and summary} \label{sint}

With the help of radial quantization, local operators in a 
conformal field theory in $d$ dimensions 
(CFT$_{d}$) can be mapped in a one to one
fashion to states in the
same CFT on $S^{d-1}\times \RRR$, 
with $\RRR$ labelling the time
direction. This takes a somewhat trivial form in
$d=1$.
Since $S^0$ is a collection of two points, the
states live in the Hilbert space $\HH\otimes \HH$ of
two copies of the CFT$_1$. 
On the other hand
the absence of spatial separation makes
all operators in the Hilbert space $\HH$ 
of a single copy
of the CFT$_1$ local. Thus the
state operator correspondence reduces to the
standard map between operators $\hat M$
in $\HH$
and states $\la a|\hat M|b\ra |a\rangle\otimes |b\rangle$
in the tensor product of two copies of
$\HH$.
In particular
the identity operator gets mapped to the maximally
entangled state 
$|a\rangle\otimes |a\rangle$ between the two copies of 
$\HH$.

This picture takes a geometric form for a class of
CFT$_1$ which are dual to string theory on
AdS$_2\times {\rm K}$ 
for some compact manifold ${\rm K}$.
These geometries typically arise
as
the near horizon geometries of black holes in the
extremal limit\cite{9803231,9809027}. 
In this case ${\rm K}$ contains the
 compactification manifold as well as the
 angular coordinates of the asymptotic space-time.
When we represent global AdS$_2$ as an infinite strip,
the two copies of the CFT$_1$ live on the two boundaries
of the strip. Furthermore
as argued in \cite{0809.3304,1008.3801} and
reviewed in \S\ref{scft}, a single copy of
the dual CFT$_1$
just consists of a finite number ($N$) of degenerate states
representing the ground states of the black hole in a
given charge sector. 
Thus two copies of the CFT$_1$ living on the two
boundaries of AdS$_2$ will contain $N^2$ states.
By AdS/CFT correspondence\cite{9711200} 
we  expect the 
dual string
theory on global AdS$_2$ to also contain $N^2$ states.
One of these states is easy to identify --
the Hartle-Hawking vacuum of string theory on
AdS$_2\times {\rm K}$\cite{hartle}. This is dual to
the identity operator in CFT$_1$ and hence represents
the maximally entangled state between the two copies
of the CFT$_1$.
Our goal in this 
note will be to identify possible origin of the
other states in string theory on AdS$_2\times {\rm K}$ which
are expected to exist according to the AdS$_2$/CFT$_1$
correspondence.

It is generally expected that AdS$_2$ cannot support
any finite energy excitation since this will destroy
the asymptotic boundary 
condition\cite{9812073}.\footnote{This
argument assumes that ${\rm K}$ is compact.
If ${\rm K}$ contains a non-compact piece {\it e.g.} 
$\RRR^2$, 
then there is no gap in the spectrum and hence in the
infrared limit we can get finite energy excitations. 
We can use local fields
to generate the corresponding states in string theory
on  in AdS$_2\times {\rm K}$, leading to
non-trivial 
correlators\cite{0907.2694,0911.3586,1001.5049}.} 
This is not a problem for us since in 
CFT$_1$ all states are of the same
energy (which we can take to be zero by a shift) and
hence we need to look only for zero energy
excitations in AdS$_2$. 
However this rules out following the usual procedure for
constructing excitations in AdS$_2$ using
local fields in the bulk\cite{9802109,9802150} 
since this
typically produces finite energy 
excitations.
Some suggestions for constructing zero energy excitations
in AdS$_2$ were made in \cite{9812073}. However
the fragmented geometries of the type discussed in
\cite{9812073} will be absent if the charge
carried by the black hole is primitive, since this prevents
the total flux to be split into multiple aligned
fluxes each through one AdS$_2$
throat.
This still leaves open the 
possibility of contribution
from the scaling solutions described 
in \cite{0005049,0504221,0608217,0702146} involving three
or more throats, with fluxes through different throats aligned
along different directions in the charge lattice.
But given that the phase space associated with these
configurations has finite volume
preventing the centers to come arbitrarily close to each
other in the quantum 
theory\cite{0807.4556,0906.0011,appear}, 
it is more 
natural to count
their effect as part of multi-centered black holes rather
than as part of a single AdS$_2$ throat. In any case
in $\NN=4$ supersymmetric string theories there is
reasonable evidence that solutions with 
three or more centers do
not contribute to the index\cite{0707.1563,0903.2481},
and hence we must look for different states.

To look for clues for where the zero energy states might
come from, let us examine the state operator correspondence
in the dual CFT$_1$. A linearly independent basis of
operators
in the CFT$_1$ is provided by the set of all
$N\times N$ hermitian
matrices.
We shall find it more convenient to work with
$N\times N$
unitary matrices instead; if we have sufficiently large number
of these matrices then 
any other matrix can be expressed as
linear combinations of these matrices.
The correlation functions in CFT$_1$
on $S^1$
are then traces of products of these matrices.
Furthermore since all the $N$ 
states in CFT$_1$ are
degenerate these $U(N)$ transformations generate
exact symmetries of the theory.
By AdS$_2$/CFT$_1$ correspondence this
symmetry must be present in the dual
string theory as well.
Thus
to compute these correlation functions in the
dual string theory on AdS$_2\times {\rm K}$ 
we represent 
euclidean global
AdS$_2$ as a disk so that the boundary on which CFT$_1$
lives becomes a circle, and then
compute a $U(N)$
twisted partition function 
in which we require the fields to satisfy a
twisted boundary condition along the boundary of
AdS$_2$\cite{0911.1563,1002.3857}, the twist being
related to the product of the matrices in CFT$_1$
whose correlation function we wish to compute. 
This suggests that 
when we represent AdS$_2$ as a strip,
we can construct the
states in string theory on AdS$_2\times {\rm K}$ 
via euclidean path integral
as in the case of
Hartle-Hawking state,
albeit  with a
twisted boundary condition in the
asymptotic past. This way
the matrix elements between these states 
naturally produces the
twisted partition function. 

Formally this prescription 
gives a complete map between
the CFT$_1$ operators and correlation functions and
the corresponding quantities in string theory 
on AdS$_2\times {\rm K}$. 
The main
problem of realizing this idea is that at
present we do not know of any explicit construction
of such $U(N)$ symmetries in string theory on 
AdS$_2\times {\rm K}$. 
However there are special
cases where we can realize a small part of this
symmetry. Typically as we move around in the moduli
spaces of a supersymmetric string theory, we encounter
special points at which there are enhanced discrete
symmetries (not to be confused with enhanced
continuous symmetries). Since typically
the black hole microstates
get transformed into each other under this discrete
symmetry, this has a
non-trivial embedding in $U(N)$. 
The dual string theory on AdS$_2\times {\rm K}$ 
also has this
symmetry manifest and we can use this to construct
the twisted states in AdS$_2$.
While this is far from providing a complete 
construction of all the states of string theory on
AdS$_2\times {\rm K}$, 
this at least
demonstrates that it is possible to construct non-trivial
states in AdS$_2$ without destroying the asymptotic
boundary conditions.
 To this end we note that
even if the near horizon geometry possesses an enhanced
discrete symmetry, it need not be a symmetry of the
asymptotic theory where the moduli can take
different values. Thus our ability to construct these
special states is not tied to the existence of some symmetry
at infinity that allows us to distinguish different black holes
trivially by doing appropriate scattering experiments
{\it e.g.} in \cite{wil1,wil2}.

This picture also incorporates naturally the dual
interpretation of the entropy of an extremal black hole.
It has been known since the work of
Bekenstein and Hawking that black holes carry entropy.
One natural explanation of this entropy is that a single
black hole represents a large collection of quantum states,
and the black hole entropy is given by the logarithm of the
degeneracy of  microstates the black hole represents. 
Indeed one of the major successes
of string theory has been to reproduce the black hole entropy
from the counting of states in the microscopic description
of the black hole\cite{9601029,9602065}. 
On the other hand the geometry of the
black hole, which includes a horizon, suggests an
alternate interpretation: the black hole entropy represents
the result of entanglement between the degrees of freedom
living outside the horizon and the degrees of freedom 
living inside the 
horizon\cite{bombelli,9303048,9401070,9403137,9404039,
0106112,0002145,0603081,0508217,0606205,
0704.0140,0709.0163,0710.2956}.\footnote{For a different
viewpoint on the relationship between black hole
entropy and entanglement see \cite{1101.3559} 
and references
therein.}
In the framework of AdS$_2$/CFT$_1$ correspondence
we see that both interpretations are equally good.
The black hole entropy $\ln N$ can be interpreted as
the logarithm of the degeneracy of a single copy of the
CFT$_1$ living on one of the boundaries of AdS$_2$, 
or as the entanglement entropy between
the two copies of CFT$_1$ living on the two boundaries
of AdS$_2$ in the Hartle-Hawking vacuum.
Since the latter corresponds to
a maximally entangled state, its entanglement entropy is
given by $\ln N$.

This observation of course
is not new -- it is the zero temperature verson of the
well known connection between black holes and
thermofield dynamics. Given any thermal
system, there is a standard doubling trick that allows us to
express the thermal averages as quantum mechanical 
expectation values in an auxiliary system containing two copies
of the original Hilbert space\cite{thermo}, and the thermal
entropy of the original system can be regarded as the
entanglement entropy of the auxiliary system.
This correspondence was exploited in 
\cite{israel,9805207,9808017,0105219,0106112} to
identify the two copies of the Hilbert space as being associated
with the two boundaries of the extended space-time for a
black hole solution.
In a related development 
it was observed in \cite{0710.2956} that if we take the
global AdS$_2$ space-time that arises in the near horizon
geometry of a black hole in the extremal limit, and calculate
the entanglement entropy between the quantum theories
living on the two boundaries of this
global AdS$_2$, then in the classical limit
the result agrees with the Wald entropy.
The argument given above shows that this must
continue to hold in the full quantum theory. While we
have a prescription for computing the degeneracy of
states in the full quantum theory as a partition function
of string theory in AdS$_2$\cite{0809.3304}, 
the prescription of \cite{0710.2956} for the
holographic computation of the
entanglement entropy in CFT$_1$ involves
evaluating the partition function of string theory
on a space-time with
conical defect.  
At the classical level the two entropies calculated using these
two apparently different computations give the same
result, but 
it is not 
clear that this equality will
continue to hold in the full quantum
theory. 
In \S\ref{sentangle} we suggest a different approach to
computing the entanglement entropy of CFT$_1$
holographically that does not entail any conical defect
and makes the equality of statistical and entanglement
entropy manifest even in the quantum theory.

\sectiono{CFT$_1$ and its state operator 
correspondence} \label{scft}

We shall begin by reviewing the properties
of the 
CFT$_1$ dual to string theory
on AdS$_2\times {\rm K}$ that arises as the near horizon
geometry of some extremal black hole.
By the usual rules of AdS/CFT correspondence
this CFT$_1$
must be given by the infrared
limit of the quantum mechanics describing the dynamics
of the brane system producing the black hole. 
In
known examples, {\it e.g.} the D1-D5-p system producing
a five dimensional black hole\cite{9601029,9602065}, 
or the 
D1-D5-p-KK monopole
system producing a four dimensional black 
hole\cite{9607026,0412287,0505094,0605210}, 
the spectrum
of the underlying 
quantum system has a gap separating the BPS ground
states from the first excited states in a fixed charge sector.
The gap is small when the charges are large, but is nevertheless
non-zero.
As a third example consider a BPS black hole in type IIB string
theory compactified on a Calabi-Yau 3-fold $CY_3$, 
described as a configuration of 3-brane wrapped on
an appropriate supersymmetric three cycle inside $CY_3$. 
The quantum mechanics
describing the system is a (0+1) dimensional sigma model with
the moduli space of supersymmetric 3-cycles as target space.
Again as long this moduli space is compact we expect the
spectrum of the quantum theory to be discrete, and there will
be a gap between the supersymmetric
ground states and the first excited state.
We shall assume that this is always the case for the quantum system
describing an extremal black hole.
Then in the infrared limit only the ground states of this
quantum mechanics will survive, and
the CFT$_1$ will consist of a finite number $N$
of degenerate
states.

The usual state operator correspondence in a $d$ dimensional
conformal field theory
relates every local operator in the conformal field theory 
to a state in the conformal field theory on 
$S^{d-1}\times \RRR$.
For $d\ge 2$ this is usually achieved by the standard
map 
from $S^d$ to $S^{d-1}\times \RRR$  via 
the coordinate
transformation that takes the north and the south poles
of $S^d$ to $\pm\infty$ of $\RRR$. 
In this case local operators inserted at the south pole
of $S^d$ create the corresponding states at $\tau=-\infty$
on $S^{d-1}\times \RRR$. 
The state operator
correspondence in $d=1$ works in a more or less similar
way. 
First the map from $S^1$ to $S^0\times \RRR$ is 
achieved via the coordinate transformation
\be \label{ens5.5a}
\sigma+i\tau
=2\tan^{-1}\tanh\left(i\theta\over 2\right)\, .
\ee
Indeed this takes the circle labelled by 
$\theta$ to a pair of lines 
$S^0\times \RRR$ where
$S^0$ corresponds to the pair of points $\sigma=0,-\pi$ and $\RRR$
is labeled by $\tau$. The points $\theta=\pm \pi/2$
are mapped to $\tau=\pm\infty$, the segment $-\pi/2<\theta
<\pi/2$ is mapped to the line at $\sigma=0$ and the
segment $\pi/2<\theta<3\pi/2$ is mapped to the line at
$\sigma=-\pi$. 
Thus CFT$_1$ on $S^0\times \RRR$ actually corresponds
to two copies of the CFT$_1$. On the other hand
since for $d=1$ there is no
notion of spatial separation, every operator acting on the Hilbert
space $\HH$ of a single copy of the
CFT$_1$ can be regarded as a local operator. Thus we are
looking for a map between the set of
operators acting on a single copy of $\HH$
to the set of states living on two copies of $\HH$
at the
two boundaries $\sigma=0,-\pi$. It is 
straightforward to construct such a
map, -- the operator $\hat M$ inserted at $\theta=
-\pi/2$ on $S^1$ creates the state
\be \label{estate}
|M\rrr = M_{ab} \, |a\rangle_{(1)}\otimes |b\rangle_{(2)}\, ,
\qquad M_{ab}\equiv \la a|\hat M|b\ra\, ,
\ee
on $S^0\times\RRR$ at $\tau=-\infty$.
Here $\{|a\rangle\}$
denotes a complete set of orthonormal
basis states in $\HH$, 
the subscripts $_{(1)}$ and $_{(2)}$ denote
the two copies of $\HH$, and
$|~\rrr$ denotes a state in $\HH\otimes \HH$.
For this state the density matrix in the Hilbert space
of the first copy of CFT$_1$, 
obtained by tracing over the states
in the second copy, is given by
\be \label{ekk1}
(M M^\dagger)_{ac} \, |a\ra \la c|\, .
\ee
Given two such states $|M\rrr$ and $|P\rrr$, we have:
\be \label{est2}
\lll M|P\rrr = M^*_{ab} P_{ab}= Tr (M^\dagger P)\, .
\ee
This can be interpreted as 
the two point function of $\hat M^\dagger$ and $\hat P$ 
in CFT$_1$ on $S^1$, in accordance with the usual
rules
of state operator correspondence in conformal
field theories.

A special state corresponding to the identity operator
in the CFT$_1$ is
\be \label{est3}
|I\rrr = |a\rangle_{(1)} \otimes |a\rangle_{(2)}\, .
\ee
We shall refer to this state as the vacuum state although 
all
states have equal energy. The corresponding density matrix
is $|a\rangle \la a|$, showing that it is a maximally entangled
state.  This however is not the only maximally entangled
state, -- it follows from \eqref{ekk1} that for any unitary
operator $\hat W$ 
the corresponding state $|W\rrr$ 
has density matrix $|a\ra \la a|$, and hence describes
a maximally entangled state. Furthermore \eqref{est2}
shows that
for unitary operators $W$ and $V$, 
the inner
product $\lll W| V\rrr$ is given by $Tr( W^{-1} V)$.

Note that in CFT$_1$, $Tr(W)$ may be expressed as
\be \label{es4}
\lll I|W_{(1)}|I\rrr\, ,
\ee
where $W_{(1)}$ denotes the operator $W$ acting on the first
copy of the Hilbert space. 
Thus CFT$_1$ correlation functions can be interpreted as
the expectation values of the operators acting on the first
copy of the CFT$_1$ in the vacuum state.

Our goal will be to seek possible representation of these
states in dual string theory on AdS$_2\times {\rm K}$.

\sectiono{AdS$_2$ space in different coordinates} \label{sgeom}

\begin{figure}
\centerline{\includegraphics[totalheight=5cm]{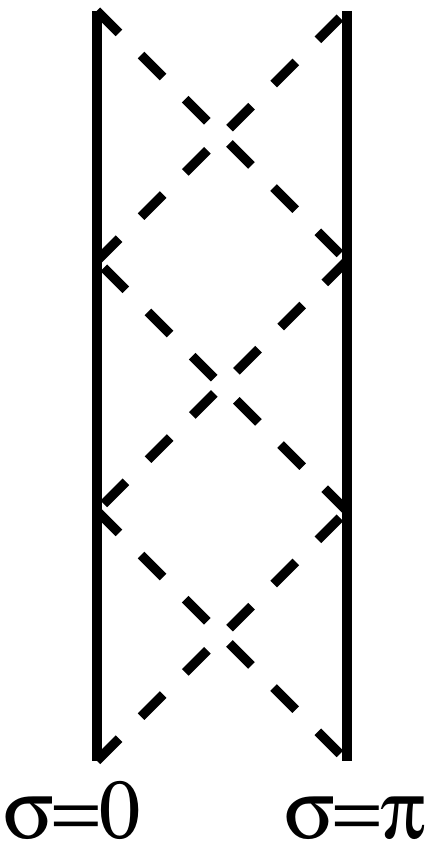}}
\caption{Global AdS$_2$ and the location of the horizon(s).
The two vertical solid lines label the two boundaries of
AdS$_2$ at $\sigma=-\pi$ (left) and $\sigma=0$ (right).
The dashed lines label the locations of the event horizons
of the original black hole.}
\label{f1}
\end{figure}

In this section we shall review some facts about the
near horizon geometry of black holes in the extremal limit.
For higher dimensional branes one usually takes
a brane solution at zero temperature and
then takes the near horizon
limit to get an AdS space-time. 
This corresponds to looking at excitations whose
energies are small from the point of view of the
asymptotic observer but large compared to the
temperature of the brane. This is
not a sensible limit for a black hole since,
as reviewed in \S\ref{scft}, the black hole 
quantum mechanics
has a gap that separates the ground state from the
first excited state, and so the only low energy
excitations are zero energy excitations.
So the sensible infrared limit is to take
the energy scale to zero as we take the temperature
to zero.\footnote{I wish to thank Hong Liu for a
discussion on this point.}
This can be achieved by taking the extremal limit in an 
appropriately rescaled
coordinate system in which the two horizons remain
finite coordinate distance away from each 
other\cite{0805.0095,0809.3304}. In this limit
part of the near horizon geometry of the black
hole involving the time and the radial coordinates
takes the form\cite{9707015,9709064,9812073}
\be \label{ens4}
ds^2 = a^2\left[ -(r^2-1) dt^2 + {dr^2 \over r^2 -1}\right]
\, ,
\ee
where $a$ is some constant.
Here,  up to a rescaling,
$r$ and $t$ can be identified as the radial and the time
variables of the full black hole solution.
The inner and the outer horizons are at $r=\pm 1$. 
The metric \refb{ens4} describes a locally AdS$_2$ space-time.
This 
can be extended to
global AdS$_2$ with the help of the coordinate 
transformation\cite{9812073}:
\be \label{ens5}
T\pm\sigma = 2\tan^{-1} \tanh{1\over 2} \left(t \pm
{1\over 2} \ln {r-1\over r+1}\right)\, .
\ee
In this coordinate system the metric takes the form:
\be \label{ens5.1}
ds^2 = {a^2\over \sin^2 \sigma} (-dT^2 + d\sigma^2)\, .
\ee
The range of $(T,\sigma)$ can be taken to be 
$(-\pi<\sigma<0$, $-\infty<T<\infty$).
This space has two boundaries, at $\sigma=0$ and at
$\sigma=-\pi$. These two boundaries lie on opposite sides
of the horizon $r=\pm 1$ of the original metric \refb{ens4}.
The asymptotic boundary $r\to\infty$ in the original metric
\refb{ens4} lies at $\sigma=0$. Fig.\ref{f1} shows AdS$_2$
in the $(\sigma,T)$ coordinate system where the locations
of the horizons at $r=\pm 1$
have been shown by the dashed line\cite{9812073,9904143}. 

Let us now consider the euclidean version of the metrics
\refb{ens4} and \refb{ens5.1}. The euclidean version
of the metric \refb{ens4} is obtained by replacing $t$
by $-i\theta$. This gives the metric
\be \label{ens5.2}
ds^2 = a^2\left[ (r^2-1) d\theta^2 + {dr^2 \over r^2 -1}\right]
\, .
\ee
Introducing new coordinate $\rho=\sqrt{(r-1)/(r+1)}$ 
we can express
the metric as
\be \label{ens5.3}
ds^2 = {4\, a^2\over (1-\rho^2)^2}
 \left[d\rho^2 +\rho^2 d\theta^2\right]\, .
\ee
In this coordinate it is clear that absence of conical singularity
at $\rho=0$ ($r=1$) requires $\theta$ to be a periodic coordinate
with period $2\pi$. The resulting two dimensional 
space spanned by $(\rho,\theta)$ with $0\le \rho<1$,
$\theta\equiv \theta+2\pi$ describes a unit disk.

Euclidean version of the metric \refb{ens5.1} is obtained
by replacing $T$ by $-i\tau$. This gives
\be \label{ens5.4}
ds^2 = {a^2\over \sin^2 \sigma} (d\tau^2 + d\sigma^2)\, .
\ee
This time there is no periodicity requirement of $\tau$.
This describes an infinite strip spanned by $(\tau,\sigma)$
with $-\infty<\tau<\infty$, $0< \sigma< \pi$.

\begin{figure}
\centerline{\includegraphics[totalheight=5cm]{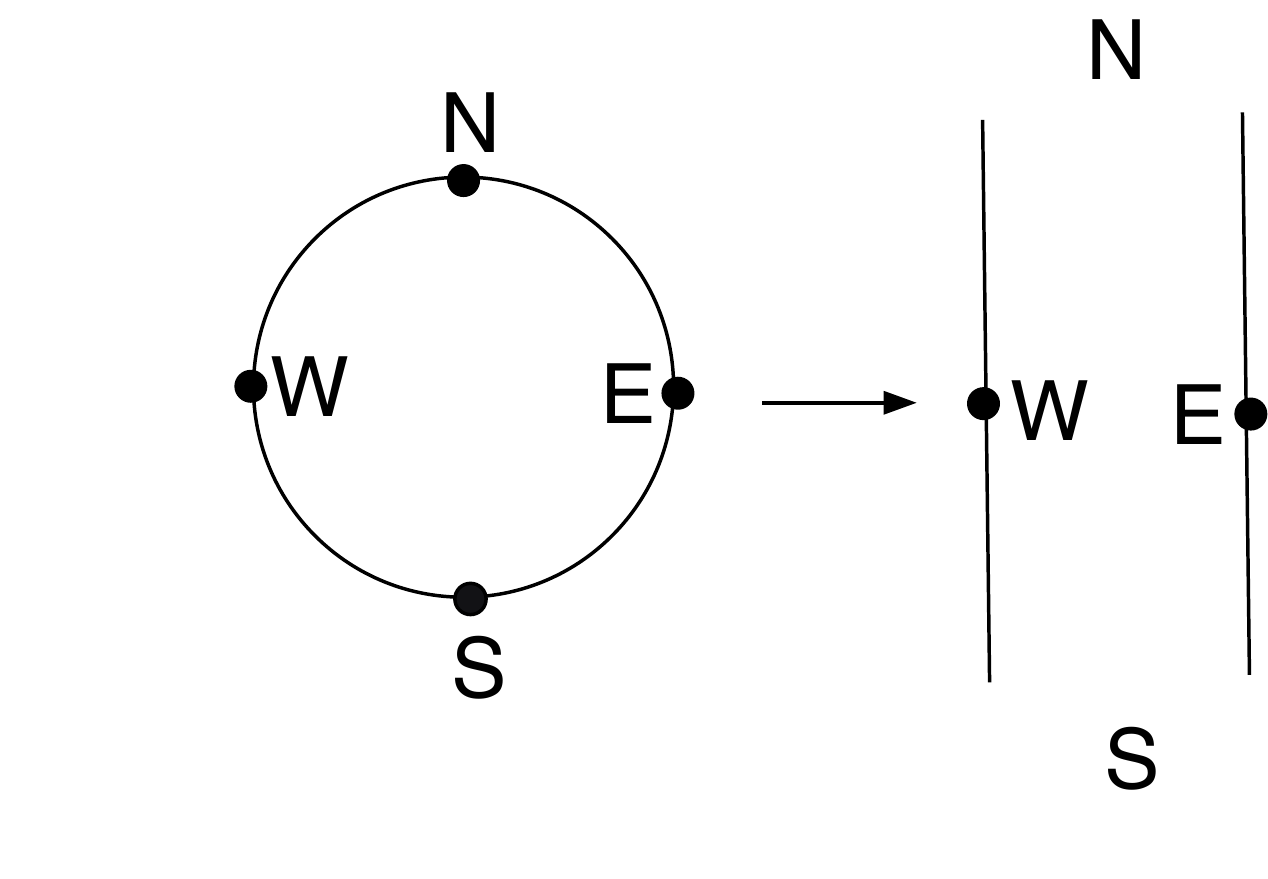}}
\caption{Conformal map from unit disk to the strip.
The left boundary of the strip is at $\sigma=-\pi$ and the
right boundary is at $\sigma=0$.}
\label{f3}
\end{figure}

Even though for the Lorentzian signature
the coordinates $(r,t)$ for $r\ge 1$ cover
only a patch of the global AdS$_2$ spanned by the
coordinates $(T,\sigma)$ in \refb{ens5.1}, the
euclidean spaces \refb{ens5.2} and \refb{ens5.4} have
an exact one to one map:
\be \label{ens5.5}
\sigma+i\tau = 
2\tan^{-1} \tanh{1\over 2} \left(
{1\over 2} \ln {r-1\over r+1}+ i\theta\right)
=2\tan^{-1}\tanh{1\over 2} \left(\ln\rho+i\theta\right)\, .
\ee
This is the standard one to one
conformal map between the unit disk
and the infinite strip as shown in Fig.~\ref{f3}. 
The segment of the boundary of the
disk $\left(\rho=1,
-{\pi\over 2}< \theta <{\pi\over 2}\right)$ gets maps to the
$\sigma=0$ boundary of the strip, and the segment 
$\left(\rho=1,
{\pi\over 2}< \theta <{3\pi\over 2}\right)$ gets mapped to the
$\sigma=-\pi$ boundary of the strip.
In fact for $\rho=1$ this reduces to the standard map from
$S^1$ to $S^0\times \RRR$  described in 
\refb{ens5.5a}.

\sectiono{AdS$_2$/CFT$_1$ correspondence} \label{scorr}

Next we shall review some aspects of the 
AdS$_2$/CFT$_1$ 
correspondence proposed in \cite{0809.3304}. Consider an
AdS$_2\times {\rm K}$ geometry 
arising as the near horizon limit of an extremal black
hole carrying a fixed charge. 
Then this is dual to the CFT$_1$ obtained as the infrared
limit of the brane system describing the dynamics of the
black hole. If CFT$_1$ has $N$ states, then
the black hole entropy, identified
as the logarithm of the number of states of the CFT$_1$, 
is given by $\ln N$.\footnote{If the black hole solution
has hair modes then we must remove their contribution
while counting $N$\cite{0901.0359,0907.0593}.}

An algorithm for computing this entropy using
the bulk description was given in 
\cite{0809.3304,1008.3801}. 
For this we consider the euclidean AdS$_2\times {\rm K}$ geometry
given in \refb{ens5.2} and denote 
by $\wh Z_{AdS_2}$ the partition function of string theory
in AdS$_2\times {\rm K}$, computed with the natural
boundary condition
that requires us to fix the electric fields at infinity
and integrate over the $r$ independent 
modes of the gauge fields.\footnote{As discussed in
\cite{0809.3304}, this requires introducing a Wilson
loop operator along the boundary of AdS$_2$ while
computing $\wh Z_{AdS_2}$.}
Due to infinite size of the
euclidean AdS$_2$ space this partition function is divergent,
so we need to regularize the divergence by putting a
cut-off on $r$, say $r\le r_0$ or equivalently
$\rho\le 1-\eps$. This makes  AdS$_2$ have
a finite volume and the boundary of AdS$_2$, situated
at $r=r_0$, have a finite length  which
we shall call $L$. Now by AdS/CFT correspondence 
$\wh Z_{AdS_2}$ should be given by the partition function of the
CFT$_1$ living on the boundary circle at $r=r_0$.
The latter in turn
is given by $Tr e^{-LH} = N \, e^{-L E_0}$, where $H$ is the
Hamiltonian of CFT$_1$ and 
$E_0$ is the energy of the $N$ degenerate states
of CFT$_1$. Thus we have
\be \label{ew1}
\wh Z_{AdS_2} = N \, e^{-E_0 L}\, .
\ee
This suggests that in order to calculate $N$, we first calculate
$\wh Z_{AdS_2}$ and then extract its finite part by expressing
it as $d_{hor}\, e^{CL}$ for some finite constants $d_{hor}$ and
$C$ in the $L\to\infty$ limit. In that case $C$ can be identified with
$-E_0$ and $d_{hor}$, called the quantum entropy function, can
be identified as the ground state degeneracy $N$ of the
black hole. 
This gives a complete prescription for computing the
black hole entropy in the bulk theory.
In the classical limit $d_{hor}$ defined this way
reproduces
the exponential of the Wald entropy\cite{0809.3304}.
In principle quantum corrections may be computed 
directly\cite{0810.3472,0904.4253,1005.3044}, 
or, for supersymmetric
black holes, using localization\cite{0905.2686}.
Significant advances towards computing $d_{hor}$ using
localization techniques have been made recently
\cite{1012.0265}.

By 
adjusting the
boundary terms in the action describing string theory
on AdS$_2\times {\rm K}$ we can make the constant $C$
vanish, so that $\wh Z_{AdS_2}$ can be directly identified as the
degeneracy of CFT$_1$. This corresponds to a constant shift
in the definition of the energy of CFT$_1$ to make $E_0$ 
vanish. For simplifying the notation we shall proceed
with this convention although we can always include
the explicit $E_0$ dependence in all the equations below
if so desired.

Can we use the bulk description to calculate 
other observables in CFT$_1$? Since CFT$_1$ consists
of a finite number of degenerate states, the only observables
are $N\times N$ matrices $M$ 
acting on this $N$
dimensional vector space. 
Let us focus on the cases of unitary matrices
which 
generate $U(N)$ transformations
in this $N$ dimensional vector space. 
Since the all $N$ states in CFT$_1$ are degenerate,
$U(N)$ is an exact symmetry of CFT$_1$. Hence it must
also exist as an exact symmetry of the dual string theory on
AdS$_2\times {\rm K}$, and corresponding to any $U(N)$
element $W$ there must be a corresponding transformation
(also denoted by $W$)
acting on the variables in the dual string 
theory.\footnote{In AdS/CFT correspondence 
global
symmetries in the boundary theory arise as local
symmetries of the bulk theory. We shall not try to
make this distinction here since we shall use the $U(N)$
transformations to twist the boundary condition, and
for this only the global part of the  group is
relevant anyway. However it is important that
there should not be any dynamical $U(N)$
gauge
fields in the bulk since this will force the black hole to
carry fixed charges under the Cartan generators
of $U(N)$\cite{0809.3304}.}  
Computing
$Tr(W)$ in CFT$_1$ will then correspond to evaluating the
partition function of string theory
on AdS$_2\times {\rm K}$ with a $W$ twisted boundary
condition on the bulk fields under $\theta\to\theta+2\pi$. 

While this gives a way to relate 
$Tr(W)$ in the boundary theory
to a specific quantity in the bulk theory,
in general the action of $W$ on the bulk fields is not
known. But in special cases, {\it e.g.} when $W$
represents a known 
$\ZZZ_k$ symmetry generator of the theory,
there is a natural lift of the action of $W$ to the bulk 
fields. In this
case $Tr(W)$ in CFT$_1$ can be identified as a
$\ZZZ_k$ twisted partition function in the bulk theory.
Since the boundary circle of the euclidean AdS$_2$
space described by the
metric \refb{ens5.3} is contractible 
in the interior,
a $W$ twisted boundary condition is not allowed there.
Thus the original AdS$_2\times {\rm K}$ geometry
does not contribute to this amplitude. But a $\ZZZ_k$
orbifold of the AdS$_2\times {\rm K}$ geometry does
contribute and gives a non-zero answer for 
$Tr(W)$\cite{0911.1563}.
This prescription has passed non-trivial tests in a class
of $\NN=4$ supersymmetric string theories where an independent
microscopic computation of $Tr (W)$ is  
possible\cite{0911.1563,1002.3857}.\footnote{Although 
the
computation of $Tr(W)$ in the bulk theory will be
the same as the one described in \cite{0911.1563}, the
spirit in which we want to use this is different. In
\cite{0911.1563,1002.3857} the asymptotic moduli 
were adjusted to also
have the unbroken $\ZZZ_k$ symmetry so that we
could compute $Tr(W)$ microscopically and compare with
the macroscopic result. In contrast, here we
want to interprete $Tr(W)$ as an observable in
the near horizon theory irrespective of whether or not
it is a symmetry of the asymptotic theory. For this we only
need to adjust the asymptotic moduli to be in a certain
subspace which under attractor flow\cite{9508072,
9602111,9602136} approaches
the point of enhanced discrete symmetry at
the horizon. \label{fo1}}
Note that
for $W=1$ we recover the original partition function 
$\wh Z_{AdS_2}$ on the
bulk side and $Tr(1)=N$ on the CFT$_1$ side.

\begin{figure}
\centerline{\includegraphics[totalheight=5cm]{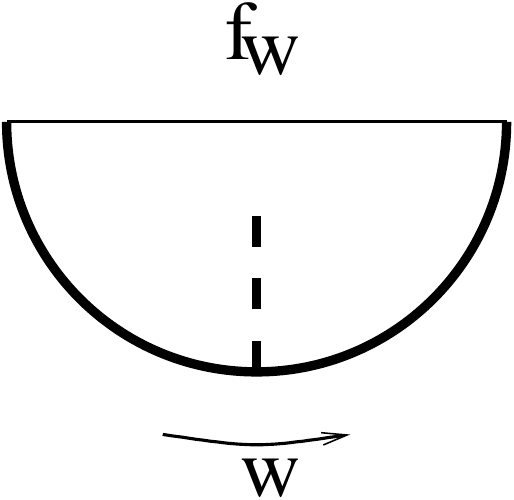}}
\caption{Generating a state in string theory on AdS$_2$
from a state $W_{ab}|a\rangle_{(1)} |b\rangle_{(2)}$
in the two copies of the Hilbert
space of CFT$_1$. The thick semi-circular line is the boundary
of AdS$_2$ whereas the thin diameter is the line on which the
string fields, appearing in the argument of $f_W$, live. The
dashed line reaching the boundary of AdS$_2$ denotes a cut
which relates the field configurations on the right of the
cut to those on the left of the cut by a transformation by
$W$.}
\label{f2}
\end{figure}

\sectiono{States in string theory on AdS$_2$}
\label{sadsstates}

The result of the previous sections
suggests a way of associating the states in two
copies of CFT$_1$ with
states in string theory on AdS$_2\times {\rm K}$.
Let us for definiteness work with states in CFT$_1$ on
$S^0\times \RRR$ of the form $|W\rrr$ with 
unitary operators $W$. 
To the states
$|W\rrr$ and $|V\rrr$ 
we want to associate wave-functions $f_W$ and $f_V$ in 
string
theory such that the inner product of $f_W$ and $f_V$ generates
the CFT$_1$ two point function $Tr(W^{-1}V)$. The latter
in turn is described by the
path integral of string theory
in AdS$_2\times {\rm K}$, with the boundary condition
that as $\theta\to\theta + 2\pi$ near the boundary, the fields are
twisted by $W^{-1}V$. This can be achieved by defining $f_W$
to be generated by the result of string theory path integral
over the half disk (or semi-infinite strip) with a cut corresponding
to the transformation $W$ that 
reaches the boundary of the half
disk (see Fig.~\ref{f2}).
The inner product $\lll W|V\rrr$ will then be obtained by gluing
the two half disks, with cuts $W$ and $V$ along the boundaries,
along their common diameter. This is given by the path integral
over the whole disk with a twisted boundary condition by $W^{-1}V$
along the boundary, as required.

Note that in Fig.~\ref{f2}
the cut can either end in the interior
of the half disk
if it is an allowed configuration in string theory,
or reach the diameter on 
which the wave-function $f_W$
is defined. This is reminiscent  of the sum over geometries
in \cite{0106112}. The necessity for allowing the cut
to reach the diameter can be seen in the computation of
$\langle\langle W|W\rangle\rangle$. The dominant 
contribution to this amplitude somes from the configuration
where the cut extends all the way across the disk,
representing the usual AdS$_2\times {\rm K}$ geometry without
any twist.

While this gives an abstract prescription for associating the
states living in two copies of the Hilbert space
of CFT$_1$ to states of string
theory on AdS$_2\times {\rm K}$, in general we cannot 
explicitly construct
these states since the action of $W$ on the bulk fields is not
known. However for the special cases when $W$
can be associated with some known
$\ZZZ_k$ symmetry generator of string theory
on AdS$_2\times {\rm K}$, we can explicitly 
construct the corresponding state;
the path integral will be over all field configurations whose
boundary values jump by the action of this discrete symmetry
transformation $W$ at some point on the boundary.
Even though this is a special case, this at least demonstrates
that it is possible for global AdS$_2$ to admit non-trivial
quantum states.

It is worth emphasizing that this discrete
symmetry need not be a symmetry 
away from the horizon --  the
asymptotic moduli could be in a
configuration that breaks this symmetry
while the attractor mechanism pulls
the near horizon geometry towards a configuration
that is invariant
under this symmetry (see footnote \ref{fo1}).
From this point of view this discrete
symmetry is on the same footing as the rest of the
proposed $U(N)$ symmetry, in that this symmetry
is present in the near horizon geometry but not necessarily
present away from the horizon.

\sectiono{Conformal invariance of the correlation
functions} \label{sl2r}

AdS$_2$ space has SL(2,$\RRR$) as a global isometry.
Thus we expect the correlation functions in CFT$_1$
to be 
SL(2,$\RRR$)
invariant.
We shall now see
how this is manifest in the formalism described above.
If we take a set of $U(N)$ elements $W_1,\cdots W_n$
then their correlation function in CFT$_1$ will be given
by $Tr(W_1\cdots W_n)$. 
This can be interpreted as the $n$-point 
correlation function on the euclidean time circle,
but since all the states have zero energy the correlation
function depends only on the cyclic time ordering of the
operators and not on the explicit time coordinates where
the operators are inserted. 
In the bulk description this 
is given by a partition function in which the boundary
values of the fields are twisted by successive applications
of $W_1,\cdots W_n$. 
Again the correlation function depends on the order in
which the $W_i$ twists are applied, but not where
we put the cut corresponding to the transformation $W_i$.
Now under an SL(2,$\RRR$) transformation the boundary
circle gets mapped to itself in a one to one fashion, and the
cyclic order of any set of points on the boundary is 
preserved.
Thus the
correlation function is manifestly invariant 
under SL(2,$\RRR$)
transformation.

The full conformal group in one dimensions is in fact
much
bigger, given by the full Virasoro group Diff$(S^1)$ that
maps the circle to itself in a one to one fashion. This also
preserves the cyclic ordering of the points on the circle
and hence is a symmetry of the correlation function in
CFT$_1$. To see how this comes about in string theory
on AdS$_2\times {\rm K}$ note that we can find a
group of diffeomorphisms in AdS$_2$
satisfying the conditions:
\begin{enumerate}
\item The elements of the group approach Diff$(S^1)$
as we approach the boundary.
\item They preserve the asymptotic boundary conditions
on the metric and
various other fields in 
AdS$_2\times {\rm K}$\cite{0803.3621,0902.1385} 
(see
\cite{0903.1477} for explicit forms of these 
diffeomorphisms
near the boundary).
\item They are well defined in the interior of AdS$_2$.
\end{enumerate}
In other words there is an asymptotic symmetry
of euclidean AdS$_2$ corresponding to the group
Diff($S^1$)
even though this is not
an isometry of AdS$_2$.
Since in computing $Tr(W_1\cdots W_n)$ in the bulk
theory we integrate over all fields subject to the
asymptotic boundary conditions, the path integral
will automatically include orbits of the above 
group of transformations. 
Furthermore these transformations will not affect the
order in which the cuts are arranged along the boundary.
Thus correlations functions
computed in the bulk theory with twisted boundary conditions
will also be manifestly invariant under the full Diff($S^1$)
group.

\sectiono{Entanglement vs statistical
entropy}   \label{sentangle}

When $W$ is the identity
operator then there is no cut on the disk, 
and the corresponding
state in string theory is simply the Hartle Hawking state obtained
by string theory path integral over the half disk without any
cut. In the boundary theory this represents the 
state $|I\rrr$ defined in \refb{est3}.\footnote{This is 
precisely the structure of the entangled
state found in \cite{0710.2956} for the special case
of extremal BTZ black holes by starting with a finite
temperature system and then taking its zero temperature
limit.}
Thus the Hartle-Hawking state 
is the maximally entangled state in the two 
copies of the CFT$_1$ living on the two boundaries of global
AdS$_2$. The corresponding entanglement 
entropy is given by
\be \label{ew6}
S_{entangle} = \ln N\, ,
\ee
where $N$ is the dimension of the Hilbert space of CFT$_1$.
This
agrees with the statistical entropy defined as
the logarithm of the degeneracy of states in the CFT$_1$.
This is a
special case of the general result of \cite{0106112}, and confirms the
explicit finding of \cite{0710.2956} that the entropy of an extremal black
hole can be interpreted as the entanglement entropy between the
two copies of CFT$_1$ living on the two boundaries of the
global AdS$_2$.
In fact not only the state \refb{est3} has entanglement
entropy $\ln N$, any twisted state $|W\rangle\rangle$
introduced earlier has density matrix $W^\dagger W =I$ and
hence entanglement entropy $\ln N$.

The equality between the statistical and entanglement
entropy is obvious in the CFT$_1$.
This can also be seen in the bulk theory as follows.
The standard procedure for computing the
entanglement entropy in a CFT uses the definition:
\be \label{erho2apre}
S_{ent} = -\lim_{n\to 1} {d\over dn} {Tr(\rho^n)\over
(Tr\rho)^n} \, .
\ee
Thus for computing  $S_{ent}$ using the
bulk description we need to find a way of computing
$Tr(\rho^n)$ holographically. This can be done as follows.
Since the
CFT$_1$ lives on a pair of points and since
the state we are interested in is the Hartle-Hawking
state, we can construct this as a path integral in the CFT$_1$
on a line of length $L/2$ with boundary conditions $\phi_1$
and $\phi_2$ at the two ends labelling the
states of the two CFT$_1$'s.
Eventually we want to take $L\to\infty$ (although since
all states have the same energy the $L$ dependence is
trivial). Now to compute the unnormalized density matrix
$\rho(\phi_1, \phi_1')$  by taking the trace over states
of the second system we take another copy of this
line segment with boundary conditions 
$\phi_1'$ and $\phi_2'$
and glue the second ends of the two segments after
identifying $\phi_2$ with $\phi_2'$. This leaves us with
a line segment of length $L$ with boundary conditions
$\phi_1$ and $\phi_1'$ at the two ends. 
For computing $\rho^n$
we simply take $n$ copies of this line segment and glue
the primed end of the $i$-th segment with the unprimed
end of the $(i+1)$-th segment for $1\le i\le (n-1)$.
This gives a line segment of length $n\, L$ with boundary
conditions $\phi_1$ and $\phi_n'$ at the two ends. 
Finally
to calculate $Tr(\rho^n)$ we join the two ends of
this line segment by identifying $\phi_1$ with
$\phi_n'$ and carry out the path integral over
the fields of the CFT$_1$ on this circle.
Thus the holographic computation of $Tr(\rho^n)$ will
involve calculating the partition function of string theory
over all spaces each of which has a boundary circle
of length $nL$ and  approaches the AdS$_2$
geometry asymptotically. 
If we denote this contribution by $\wh Z_{AdS_2}(n)$ then
we have 
\be \label{eentbulk}
Tr(\rho^n) = \wh Z_{AdS_2}(n)\, .
\ee
Hence from \refb{erho2apre} we get
\be \label{erho2a}
S_{ent} =  -\lim_{n\to 1} {d\over dn} {\wh Z_{AdS_2}(n)
\over \wh Z_{AdS_2}(1)^n}\, .
\ee
$\wh Z_{AdS_2}(1)$ can be identified with the 
$\wh Z_{AdS_2}$ given in \eqref{ew1}. To
compute $\wh Z_{AdS_2}(n)$ we note that 
the leading contribution to this
partition function comes from the euclidean
AdS$_2$ geometry \eqref{ens5.2}
itself with the cut-off on $r$ adjusted
to produce the appropriate boundary length $nL$. 
Even the
full quantum contribution to the partition function will
be given by the quantum contribution to $\wh Z_{AdS_2}$
with a different infrared cut-off so that the boundary has
length $nL$.
Thus we have from \eqref{ens5.2}
\be \label{erho1}
\wh Z_{AdS_2}(n) = N e^{-n\, E_0 \, L}\, .
\ee
Substituting this into \refb{erho2a} gives
\be \label{erho2}
S_{ent} =  -\lim_{n\to 1} {d\over dn} N^{1-n}
= \ln \, N\, ,
\ee
which is manifestly equal to the statistical entropy. 

Note
that this approach differs from the holographic prescription
of \cite{0710.2956,0905.0932} 
for the computation of 
$Tr(\rho^n)$. In the latter approach while computing
$\wh Z_{AdS_2}(n)$ we compute the partition function of
of string theory on $n$-fold cover of AdS$_2$. 
This has a conical defect at the center.
On the other
hand in our approach we first take an $n$-fold cover of the
boundary circle and then integrate over all possible bulk
space-time with this boundary condition. 
The leading saddle point is the AdS$_2$ space itself with
a different infrared cut-off. Although classically the two
approaches give the same result, it is not clear {\it a
priori} if the agreement will continue to hold in quantum string
theory. Indeed at present we do 
not know how to define string
theory on a space with conical defect for an arbitrary
defect angle. Clearly understanding the relationship between
these two computations will be desirable since it might also
given us a clue as to when, why and how the holographic
prescription\cite{0710.2956,0905.0932,0603001,0605073} 
for
computing entanglement entropy works.

\sectiono{Information loss problem}
\label{sinfo}

Since an extremal black hole has zero temperature, we cannot
directly formulate the usual information loss puzzle involving
absorption and subsequent Hawking radiation from such a black hole
unless we allow the black hole to go through an intermediate
non-extremal state. 
The best we can do is the following.
Suppose we  
probe the extremal
black hole by an external agent carrying energy lower than
the gap that separates the ground state of the black
hole from the first excited state. In this case 
the only possible
transitions are to the other ground states of the
black hole. 
Will the black hole retain a memory of this probe that can be
tested using a second experiment with
another probe?
To address  this question 
we note that
the effect of such probes can be 
described by (linear combinations of) the
twist operators described earlier.\footnote{In 
general the relation
between the external probe and the twist operator will
be complicated and depend on the interpolating
geometry  that
connects the near horizon region to the asymptotic
region. For example in the case of the $\ZZZ_k$ symmetry
which represents a known discrete symmetry of string
theory at special points in the moduli
space, the relation between the twist operators and
external probes depends on how a low energy probe at infinity
where the $\ZZZ_k$ symmetry may be broken, transforms 
itself into
a linear combination of $\ZZZ_k$ eigenstates by the time it
reaches the horizon where the $\ZZZ_k$ symmetry is unbroken.
Such a relation can in principle
be derived from the
knowledge of the full black hole solution that interpolates
between the near horizon geometry with unbroken
$\ZZZ_k$ symmetry and the asymptotic region.}
If we now perform two successive experiments on the
black hole, one with the probe $W$ followed by another
with probe $V$, then the effect of the first probe on the second
experiment will be tested by
the two point function of $W$ and $V$.
Since in general this two point function is non-zero, we see
that the black hole does retain the
memory of the first probe.
The simplest example of this is the case where $WV$
represents a $\ZZZ_k$ twist of the type discussed
in \cite{0911.1563,1002.3857}. In this case the leading saddle
point that contributes to this correlation function is a
$\ZZZ_k$ orbifold of AdS$_2\times {\rm K}$, and
has a contribution of order
$N^{1/k}$\cite{0911.1563} 
compared to the contribution of order $N$ to
$Tr(1)$.
Thus
as in the case of \cite{0106112}, 
the non-vanishing contribution 
comes only after we sum over non-trivial
saddle points, and is suppressed by  a power of $N$.

\sectiono{Speculations on the enhanced symmetry}
\label{sspec}

If our proposal for the state operator correspondence
is correct, this will imply that string theory in 
the near horizon geometry of an extremal black hole
has a $U(N)$ symmetry by which we can twist the
boundary conditions on the fields. 
Since $N$ can be very large this implies a large
group of symmetries of the theory. 
This is expected to be symmetry of string theory in the
AdS$_2\times {\rm K}$ geometry but not of the string theory
in asymptotically flat space time in which the black hole
is embedded, since the $U(N)$ symmetry acting on the
$N$ degenerate states of the black hole becomes an
exact symmetry only in the infrared limit in which there
is a decoupled quantum mechanics of the $N$ degnerate
BPS states.

At present we do not have any concrete understanding
of how such enhanced symmetries could arise. We
shall end by making some random observations:
\begin{enumerate}
\item In the classical limit the
black hole entropy $\ln N$ goes to infinity. Thus if the
$U(N)$ symmetry is present in the classical limit, then it
must appear as a $U(\infty)$ symmetry which is
broken down to $U(N)$ by quantum effects. 
Since $U(\infty)$ is a symmetry of the infinite dimensional
complex grassmannians, one could wonder if grassmannians
might play a role in string theory on AdS$_2\times {\rm K}$.
Alternatively
the $U(N)$ symmetry could arise only  as
a symmetry of the
quantum theory with no classical analog.
\item Typically for a BPS black hole in a supersymmetric
theory carrying a fixed charge, 
some of the moduli scalar fields are fixed at the
horizon due to the attractor mechanism, but 
the other moduli may remain free and label the moduli
space of the near horizon geometry. As we move
around in this
moduli space, the discrete symmetries of the theory
may change, being either non-existent or a small group
of symmetries at a generic point but getting enhanced to
bigger groups at special points. 
On the other hand
the spectra of the black hole ground states
at different points in this moduli space are expected to be
isomorphic 
since they represent the BPS states carrying
a given set of charges.
It should in principle be
possible to use these isomorphisms to represent the
action of the discrete symmetries at different points in the
moduli space on the same Hilbert space. Thus all these
discrete symmetries must be embedded in the
single $U(N)$ group that acts on the $N$ degenerate
ground states of the black hole.

To take a concrete example, consider type IIA
string theory compactified on $K3\times T^2$
and take a black hole that carries only fundamental
string winding, momentum, Kaluza-Klein (KK) 
monopole and H-monopole charges associated with
the two circles of $T^2$.
Although the full moduli space of the theory is
parametrized locally by the $SO(6,22)/SO(6)\times SO(22)$
coset space, some of the moduli are fixed in the near
horizon geometry leaving behind only a locally
$SO(4,22)/SO(4)\times SO(22)$ space.
The moduli labelling this space include in particular the
metric and the 2-form fields on $K3$.
Now it has been recently speculated
in\cite{0907.1410,1004.0956,1005.5415,1006.0221,
1006.3472,1008.0954,1008.3778,1008.4924} that the 
symmetry group of
supersymmetric states in a sigma model with target
space $K3$ includes the Matthew group $M_{24}$
even though there is no known point in the $K3$
moduli space at which the corresponding
string theory has manifest $M_{24}$ symmetry.
In this case this group must also have a natural action
on the space of BPS states of the black hole described
above and
sit inside the $U(N)$ group acting on the $N$ degenerate
ground states of the black hole. 
A better understanding of why $M_{24}$ appears as a
symmetry of supersymmetric states in the sigma model
could help us realize this as an explicit symmetry of string
theory in this particular near horizon geometry. This will
still fall short of realizing the whole $U(N)$ group as a manifest
symmetry, but will help us realize a large subgroup of
$U(N)$.

For special values
of the charges the symmetry group may also include
a subgroup of the duality
group associated with compactification on $T^2$.
Note however that none of these symmetries may
be a symmetry of the asymptotic theory since they
can be broken by the expectation values of the moduli
fields at infinity.
\item Supergravity theories reduced to two dimensions
typically have a large group of continuous duality 
symmetries. Normally in the presence of charged particles
this symmetry breaks down to a discrete subgroup. However
since in the AdS$_2$ geometry there are no charged
excitations
one could wonder if these continuous duality symmetries
could play any role in building up the $U(N)$ group. In this
context it is encouraging to note that the enhanced discrete
symmetries at special points in the moduli space are
naturally embedded in this continuous duality group.
\end{enumerate}

\bigskip

{\bf Acknowledgement: } I would like to thank Rajesh
Gopakumar, Dileep Jatkar,
Hong Liu, Shiraz Minwalla, Erik Tonni, 
Sandip Trivedi and Barton Zwiebach for discussion on
various aspects of this work.
This work was supported in part by the
J. C.  Bose fellowship of the Department of
Science and Technology, India and the
project 11-R\& D-HRI-5.02-0304. I would also
like to thank the Morningstar visiting professorship
for support during a visit to Center for Theoretical Physics, 
MIT.

\end{document}